\documentclass[twocolumn,showpacs,preprintnumbers,amsmath,amssymb]{revtex4}
\usepackage{graphicx}
\usepackage{bm}

\usepackage{epsfig}
\usepackage{hyperref}
\usepackage{lineno}
\usepackage{xspace}                                          
\usepackage{color}
\newcommand{\sqsn}{\mbox{$\sqrt{s_{_{NN}}}$}\xspace}
\newcommand{\sqrts}[1]{\mbox{$\sqrt{s_{_{NN}}}$}~=~#1~GeV\xspace} 
\newcommand{\AuAu}{\mbox{Au~+~Au}\xspace} 
\newcommand{\CuCu}{\mbox{Cu~+~Cu}\xspace} 
\newcommand{\PbPb}{\mbox{Pb~+~Pb}\xspace} 

\begin{document}

\title{Experimental Results on Charge Fluctuations in Heavy-ion Collisions}
\author{D.~K.~Mishra$^1$\footnote{e-mail: dkmishra@rcf.rhic.bnl.gov}, 
P.~Garg$^2$, P.~K.~Netrakanti$^1$, L.~M.~Pant$^1$, and 
A.~K.~Mohanty$^1$\footnote{Presently at Saha Institute of Nuclear Physics, 
1/AF, Bidhan nagar, Kolkata - 700064, India}}
\affiliation{$^1$Nuclear Physics Division, Bhabha Atomic Research Center, Mumbai 
400085, India, \\
$^2$Department of Physics and Astronomy, Stony Brook University, SUNY, 
Stony Brook, New York 11794-3800, USA \\
}
%

\begin{abstract}
We present a subset of experimental results on charge fluctuation from the 
heavy-ion collisions to search for phase transition and location of critical 
point in the QCD phase diagram. Measurements from the heavy-ion experiments at 
the SPS and RHIC energies observe that total charge fluctuations increase from 
central to peripheral collisions. The net-charge fluctuations in terms of 
dynamical fluctuation measure $\nu_{(+-,dyn)}$ are studied as a function of 
collision energy (\sqsn) and centrality of the collisions.
The product of $\nu_{(+-,dyn)}$ and $\langle N_{ch} \rangle$ shows a 
monotonic decrease with collision energies, which indicates that at LHC energy 
the fluctuations have their origin in the QGP phase. The 
fluctuations in terms of higher moments of net-proton, net-electric charge and 
net-kaon have been measured for various \sqsn. Deviations are observed in both 
$S\sigma$ and $\kappa\sigma^2$ for net-proton multiplicity distributions from 
the Skellam and hadron resonance gas model for \sqsn $<$ 39 GeV. Higher moment 
results of the net-electric charge and net-kaon do not observe any significant 
non-monotonic behavior as a function of collision energy. We also discuss the 
extraction of the freeze-out parameters using particle ratios and 
experimentally measured higher moments of net-charge fluctuations. The 
extracted freeze-out parameters from experimentally measured moments and lattice 
calculations, are found to be in agreement with the results obtained from the 
fit of particle ratios to the thermal model calculations.
%
\end{abstract}
\maketitle
\section{Introduction}
\label{sec:introduction}
The main goal of the high energy heavy-ion collisions is to study the phase 
structure of the Quantum Chromodynamic (QCD) phase diagram at finite temperature 
($T$) and baryon chemical potential 
($\mu_B$)~\cite{Stephanov:1998dy,Alford:1997zt,Stephanov:1996ki,Aoki:2006we, 
Fukushima:2010bq}. Several theoretical models suggest that the QCD phase 
diagram may contain a first order phase transition line between the hadron gas 
(HG) phase and Quark-Gluon phase which ends at the critical point towards high 
$T$ and lower 
$\mu_B$~\cite{Stephanov:2004wx,Fodor:2004nz,Stephanov:1999zu,Mukherjee:2015swa,
Mukherjee:2016kyu}. 
Experimental programs have been performed at SPS and beam-energy scan (BES-I) 
program at RHIC to search for critical point and QGP-HG phase transition. In 
future, the upcoming program at RHIC (BES-II)~\cite{BES-II}, FAIR~\cite{FAIR}, 
NICA~\cite{NICA} and J-PARC~\cite{JPARC} will 
also contribute to the physics at large $\mu_B$. The location of the QCD
critical point can be explored by systematically varying $T$ and $\mu_B$. 
Experimentally, by changing the center of mass energy one can control the 
$T$ and $\mu_B$ of the system, hence enables us to scan different sectors of the 
phase diagram.

One of the most striking signatures of such a QGP-HG phase transition could be 
a strong modification in the fluctuations of specific observables measured on an
event-by-event basis~\cite{Asakawa:2000wh,Stodolsky:1995ds}. In principle, any 
observable that is not globally conserved, fluctuates. Although, most of these 
fluctuations are trivial and are of statistical origin. It is important to 
find out the dynamically relevant event-to-event fluctuation, that enables to 
the search for a possible critical point and a first order co-existence region 
in the QCD phase diagram~\cite{Adams:2003st}. Over the past two decades quite a 
number of such observables have been suggested for clarifying the 
evolution of the system formed in heavy-ion collisions. These either refer 
to the signals from the plasma that are supposed to survive the phase 
transition or to the observables that experience strong fluctuation during the 
phase transition or close to the critical point. Most commonly measured 
event-by-event fluctuations in heavy ion collision experiments are particle 
ratios ($K/\pi$, $p/\pi$ etc.), transverse energy $(E_{T})$, mean transverse 
momentum $\langle p_{T} \rangle$ and particle multiplicity 
$(N)$ fluctuations ~\cite{Abelev:2009ai,Alt:2008ab,Adcox:2002pa,Adler:2003xq}.
Predictions suggest that, enhanced multiplicity fluctuations are connected to 
the production of QGP droplets, and suppression of fluctuation is connected 
to the nucleation process in a first order QGP--HG phase transition. This may 
happen because of the rapid freeze-out just after the phase 
transition~\cite{Asakawa:2000wh,Jeon:2000wg}.
An isothermal compressibility of the system can be considered to understand the 
sensitivity of the measured particle multiplicity to the phase 
transition~\cite{Mrowczynski:1997kz}. The isothermal compressibility is defined 
as $k_T = -1/V(\delta V/\delta P)_T$, where $P, V$ and $T$ being the 
pressure, volume, and temperature of the system, respectively. In order to 
relate the compressibility to the measurements of multiplicity fluctuations, we 
assume that relativistic heavy-ion collisions can be described as a thermal 
system in the Grand Canonical Ensemble (GCE)~\cite{Begun:2004pk}. The GCE is the 
most appropriate description as only part of the particles from the system 
around mid-rapidity are measured by the experiments. The energy and conserved 
quantum numbers in this region can be exchanged with the rest of the system, 
that serves as a heat bath~\cite{Jeon:2003gk}. Several other studies have 
applied Canonical and Micro-canonical ensembles to the multiplicity 
fluctuations too~\cite{Begun:2004gs,Becattini:2005cc,Garg:2015owa}. In the GCE, 
the isothermal compressibility is directly related to the variance of the 
particle multiplicity as follows:

\begin{equation}
 \langle (N-\langle N \rangle)^2\rangle = var(N) 
 = \frac{k_BT\langle N\rangle^2}{V}k_T
\end{equation}
where $N$ is the particle multiplicity, $\langle N\rangle = \mu_N$ is the mean 
of the multiplicity distribution, and $k_B$ is the Boltzmann 
constant~\cite{Stanley}. Here, multiplicity fluctuation measurements are 
presented in terms of the scaled variance, $\omega_N$ as~\cite{Adare:2008ns}:

\begin{equation}
 \omega_N = var(N)/\mu_N = k_BT\frac{\mu_N}{V}k_T
\end{equation}

In a continuous, or second-order phase transition, the compressibility 
diverges at the critical point. Near the critical point, 
this divergence is described by a power law in the variable $\epsilon = (T-
T_c )/T_c$ , where $T_c$ is the critical temperature. Hence, the relationship
between multiplicity fluctuations and the compressibility can be exploited to 
search for a signature of critical behavior by looking for the expected 
power law scaling of the compressibility:

\begin{equation}
 k_T\propto(\frac{T-T_c}{T_c})^{-\gamma}\propto|\epsilon|^{-\gamma}
\end{equation}
where $\gamma$ is the critical exponent for isothermal 
compressibility~\cite{Stanley}. 
Recent studies~\cite{Schaefer:2006ds,Sasaki:2006ws} show the behavior of the 
quark number susceptibility, $\chi_q$, which is related to the value of the 
isothermal compressibility of the system. They predict that its value will 
increase by at least an order of magnitude close to the QCD critical point. As 
discussed above, the scaled variance is proportional to $k_T$, hence the 
measurements of charged particle multiplicity are expected to be a sensitive 
probe for critical behavior. 

If the system approaches close enough to the critical line for a long enough 
time period, then critical phenomena could be observed through the measurement 
of multiplicity fluctuations~\cite{Stephanov:1999zu}. Subsequently, it may 
also be possible to determine the critical exponents of the system. 
Observation of critical behavior in heavy ion collisions and the subsequent 
measurement of the critical exponents could determine the universality class in 
which QCD is grouped, providing essential constraints for the 
models~\cite{Adare:2008ns}.

The fluctuations of conserved quantities are predicted to be one of the most 
sensitive signals of the QGP formation and phase transition, which may provide 
complementary understanding of strong interactions, apart from other 
QGP signatures~\cite{Koch:2005vg,Asakawa:2000wh}. It has been argued that 
entropy conserving hadronization of a plasma of quarks and gluons should
produce a final state characterized by a dramatic reduction of the net-charge 
fluctuations in QGP phase as compared to that of a hadron gas. Further, 
prediction relies on the notion that quark-quark correlations can be neglected,
and hadronization of gluons produces pairs of positive and negative particles 
not contributing to the net-charge fluctuations. It has also been suggested 
that, the excitation function of conserved numbers like net-baryon, net-charge 
and net-strangeness fluctuations should show a non-monotonic behavior, as a 
possible signature of QCD critical end point 
(CEP)~\cite{Stephanov:1998dy,Stephanov:2011pb,Asakawa:2009aj}. In the 
thermodynamic limit, the correlation length ($\xi$) diverges at 
CEP~\cite{Stephanov:1998dy}. The experimentally measured moments of the 
net-baryon, net-charge, and net-strangeness distributions are related to the 
higher power of $\xi$ of the system and hence these moments can be used to look 
for signals of a phase transition and critical 
point~\cite{Ejiri:2005wq,Bazavov:2012vg}. Also, 
the comparison of experimentally measured cumulants with the lattice 
calculations enables us to extract the freeze-out parameters i.e. freeze-out 
temperature ($T_f$) and $\mu_B$ of the system produced in heavy-ion 
collisions~\cite{Borsanyi:2014ewa,Alba:2014eba}. In recent years, lots of 
efforts have been put on both theoretical and experimental fronts to study the 
fluctuation of conserved quantities.

This review is organized as follows: In the following section, we discuss the 
total charge fluctuations from various experiments. In Sec.~\ref{sec:netch}, 
the results on net-charge fluctuation are presented, which include dynamical 
fluctuation measure $\nu_{(+-,dyn)}$, higher moments of net-proton, 
net-electric charge and net-kaon fluctuations. Towards end of the 
Sec.~\ref{sec:netch}, extraction of freeze-out parameters using higher moments 
are discussed. Finally, in Sec.~\ref{sec:summary}, we summarize our 
observations.

\section{Total charge fluctuation}
\label{sec:totch}
In a thermodynamical system of strongly interacting matter formed in the 
heavy-ion collisions, the fluctuations of particle multiplicities, mean 
transverse 
momentum ($\langle p_{T} \rangle$), transverse energy ($E_T$)  and other global 
observables are related to the fundamental properties of the system, such as 
specific heat, chemical potential and compressibility. These observables either 
refer to signals from the plasma that are supposed to survive the phase 
transition or to observables that experience strong fluctuations during the 
phase transition or close to the critical point. The existence of a critical 
point at the QCD phase transition has been associated with the large 
event-by-event fluctuations of above observables. As far as observables are 
concerned, electric charge fluctuations have been measured over a wide range of 
collision energies, from CERN SPS to RHIC and LHC energies. Enhanced 
fluctuations in neutral to charged pions have been predicted as a signature of
the formation of Disoriented Chiral Condensates 
(DCC)~\cite{Aggarwal:2000aw,Rajagopal:1993ah}.
The relative fluctuation $\omega_N$ which can be extracted from experimental 
data has contributions both from statistical as well as dynamical sources. In 
order to extract the dynamical part associated with new physics from the 
observed fluctuations, one has to understand the contributions from statistical 
and other known sources. Some of the known sources of fluctuations contributing
to the observed experimental value of scaled variance ($\omega_N$) include, 
finite particle multiplicity, effect of limited acceptance of the detectors, 
impact parameter fluctuations, fluctuations in the number of primary 
collisions, effects of re-scattering of secondaries, resonance decays, and 
Bose-Einstein correlations~\cite{Heiselberg:2000fk}.
 
  \begin{figure}[htbp]
\includegraphics[width=0.5\textwidth]{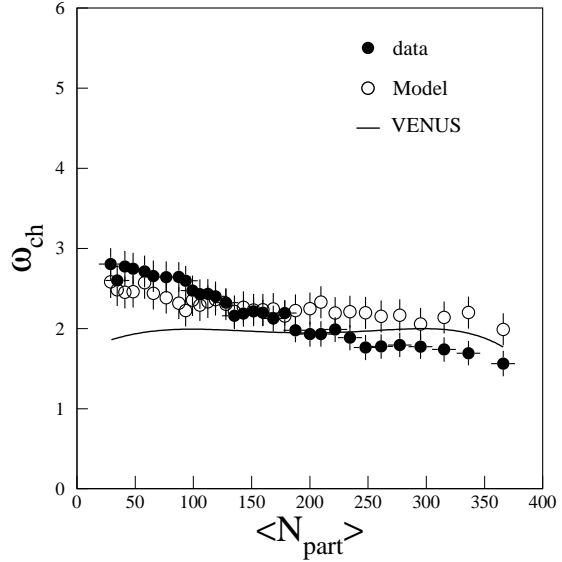}
    \caption{The relative fluctuations, $\omega_{ch}$, 
of the event-by-event measured charged particle multiplicity as a function of 
number of participants ($N_{part}$) in \PbPb collisions at 158 $A$ 
GeV~\cite{Aggarwal:2000aw}. The combined (statistical and systematic) errors on 
$\omega_{ch}$ from the experimental data are shown along with the data points. 
Details on error estimation are in Ref.~\cite{Aggarwal:2000aw}. The 
experimental data are compared to calculations from a participant model and 
those from VENUS event generator.}
\label{fig:chwa98}
  \end{figure}
  \begin{figure}[htbp]
\includegraphics[width=0.5\textwidth]{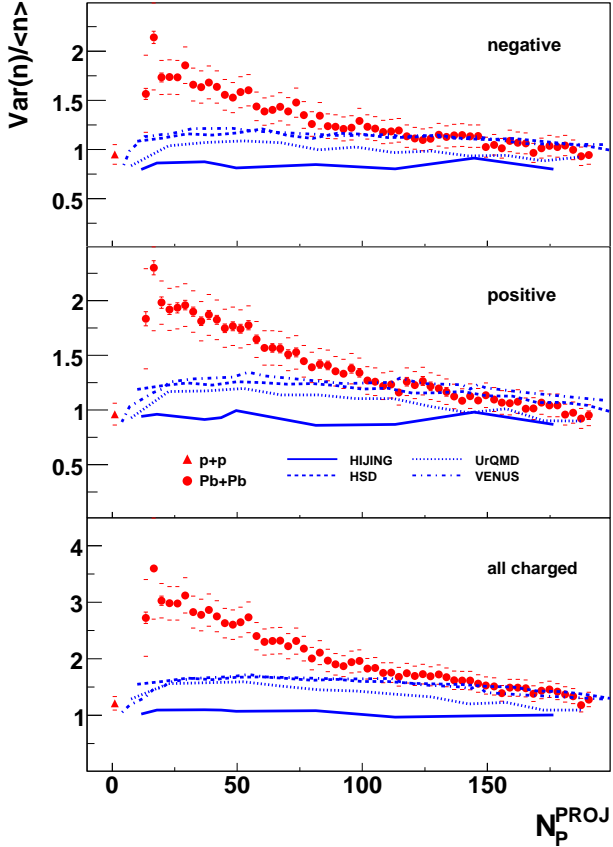}
    \caption{The scaled variance of the multiplicity
distribution for negatively (upper panel), positively (middle
panel) and total (bottom panel) charged particles as a function
of centrality in terms of number of projectile participants in \PbPb collisions 
at 158.$A$ GeV~\cite{Alt:2006jr}. The experimental results are compared with 
model simulations in the NA49 acceptance (HSD and UrQMD predictions were taken 
from~\cite{Konchakovski:2005hq}. The statistical errors are smaller than the 
symbols (except for the most peripheral points). The horizontal bars indicate 
the systematic uncertainties.}
\label{fig:chfluc_na49}
  \end{figure}

\begin{figure}[htbp]
\includegraphics[width=0.5\textwidth]{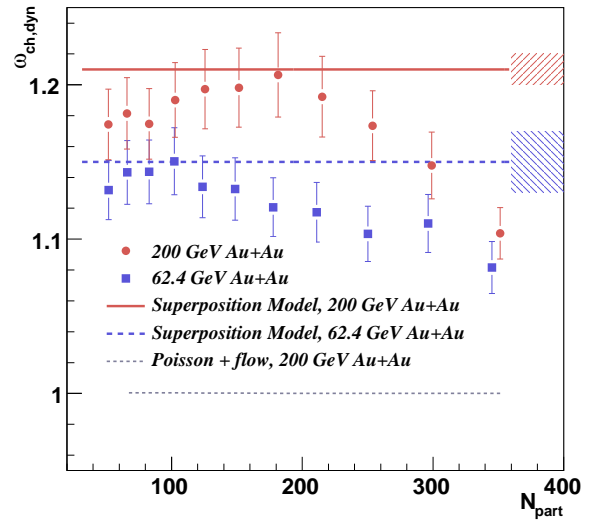}
    \caption{ Centrality dependence of the scaled variance 
in \AuAu collisions for 0.2 $< p_T <$ 2.0 GeV/c for \sqsn = 62.4 and 200 
GeV~\cite{Adare:2008ns}. 
Results from the superposition model are overladed with the shaded regions 
representing a one standard deviation range of the prediction for the 
fluctuation magnitude derived from $p+p$ collision data. Contribution from 
non-correlated particle emission with the Poisson distribution of the scaled 
variance with the addition of elliptic flow in 200 GeV \AuAu collisions are 
also shown.
}
\label{fig:chfluc_phenix}
  \end{figure}

The relative fluctuation $\omega_{ch}$ is defined as
\begin{equation}
 \omega_{ch} = \frac{\langle N_{ch}^2\rangle - \langle N_{ch}\rangle^2}{\langle 
N_{ch}\rangle} = \frac{var(N_{ch})}{\langle N_{ch}\rangle}
\end{equation}
where $N_{ch}$ is the charged particle multiplicity, $\omega_{ch}$ is also 
known as scaled variance. If the multiplicity distribution is Poissonian, the 
scaled variance is 1.0.
Figure~\ref{fig:chwa98} shows the comparison of the relative fluctuation 
$\omega_{ch}$ of the charge particle multiplicity as a function of 
collision centrality which is related to number of participants ($N_{part}$) 
in {\mbox{Pb~+~Pb}\xspace} collisions at 158 A GeV~\cite{Aggarwal:2000aw}. The 
error on $\omega_{ch}$ calculated in the model is mainly due to the error on 
the mean number of charged particles in nucleon-nucleon interactions, the error 
in the number of participants calculated, and the uncertainty in the calculated 
transverse energy~\cite{Aggarwal:2000aw}. The 
experimental data are compared with the model calculations. It is observed that 
the relative fluctuations increase from central to peripheral collisions. The 
observed charge particles multiplicity fluctuations have been found to well 
agreed with the results obtained from a simple participant 
model~\cite{Baym:1999up}. In the participant model, the particle multiplicity 
$N$ may be expressed as 
\begin{equation}
     N = \sum_{i = 1}^{N_{part}} n_i
\end{equation}
where $N_{part}$ is the number of participants in the collision and $n_i$ is 
the number of particles produced by the $i$th participant within the detector 
acceptance. The mean value of $n_i$ is the ratio of the average multiplicity 
measured in the 
detector acceptance to the average number of participants, $\langle n\rangle = 
\langle N \rangle/\langle N_{part}\rangle$. Hence, fluctuation in the particle 
multiplicity $N$ will have contributions due to fluctuations in $N_{part}$,  
($\omega_{_{N_{part}}}$) and also due to the fluctuations in the number of 
particles produced per participant ($\omega_n$). The multiplicity fluctuation in 
the participant model can be expressed as~\cite{Aggarwal:2000aw} 
\begin{equation}
 \omega_N = \omega_n + \langle n \rangle \omega_{_{N_{part}}}    
\end{equation}
Another experiment at SPS also performed similar study shown in 
Fig.~\ref{fig:chfluc_na49}~\cite{Alt:2006jr}. The scaled variance 
($Var(n)/\langle n \rangle$), where $Var(n) = (\langle n^2 \rangle - \langle n 
\rangle^2)$ is the variance of the distribution and $n$ being 
the multiplicity of the particles. The scaled variance of positive, 
negative and total charged particles as a function of centrality is shown in 
Fig.~\ref{fig:chfluc_na49}. The experimental data is compared with the model 
calculations. The results from different models (HIJING~\cite{Gyulassy:1994ew}, 
HSD~\cite{Cassing:1999es}, UrQMD~\cite{Bleicher:1999xi}, and 
VENUS~\cite{Werner:1993uh}) are almost independent of 
centrality and behave like a Poisson expectation. However, the experimental 
data points indicate strong dependence on centrality. The scaled variance 
increases from central to peripheral collisions. The measured centrality 
dependence can be reproduced in superposition models with the assumption of 
contributions from target participants to the particle production in the 
forward hemisphere~\cite{Alt:2006jr,Konchakovski:2005hq}. 
Figure~\ref{fig:chfluc_phenix} shows the 
centrality ($N_{part}$) dependence of scaled variance at \sqrts{62.4 and 200} in 
\AuAu collisions at RHIC~\cite{Adare:2008ns}. The shaded regions represent the 
systematic uncertainties from the reference range. The statistical 
uncertainties are shown along with the data points. Here, $\omega_{ch,dyn}$ 
represents the estimate of the remaining dynamical multiplicity fluctuations. 
For all centralities, the scaled variance lies above the Poisson expectation of 
1.0. At these energies, also the scaled variance increases from central to 
peripheral collisions. Hence, similar centrality dependence has been observed 
by the experiments at the SPS and RHIC energies. The absence of large dynamical 
fluctuation in excess of the participant superposition model indicates that 
there is no evidence of critical behavior related to the compressibility 
observable.

\section{Net-charge fluctuations}
\label{sec:netch}
One of the proposed signatures to search for the phase transition from hadronic 
to partonic medium is to study the net-charge fluctuations in heavy ion 
collisions. The fluctuation in the net charge depends on the square of the 
charges and hence strongly depend on which phase it originates from. The charge 
carriers in the QGP phase are quarks having fractional charges, while in 
hadronic phase the constituents have unit charge, hence the measure of the 
fluctuations in the net-charge is expected to be different 
in these two cases~\cite{Jeon:2000wg}. In this section, we discuss the 
net-charge fluctuations using different fluctuation measures such as 
$\nu_{(+-,dyn)}$ and higher moments.

\subsection{Fluctuation study using $\nu_{(+-,dyn)}$}
The net-charge fluctuations are expected to be smaller in the QGP phase as 
compared to the hadron gas (HG) phase~\cite{Jeon:2000wg}. The net-charge 
fluctuations may get affected by uncertainties due to volume fluctuation, exact 
local charge conservation or repulsive forces among hadrons~\cite{Alba:2014eba}. 
However, it is important to know whether these fluctuations may or may not 
survive the evolution of the system in the heavy-ion collisions. The collision 
volume is not directly measured in the experiment which may lead to additional 
geometrical fluctuations. One can get rid of 
volume fluctuation by considering the ratios of the number of positive ($N_+$) 
to negative ($N_-$) particles, $R = N_+/N_-$. The ratio fluctuation is defined 
by $\langle \delta R^2 \rangle = \langle R^2 \rangle - \langle R \rangle ^2$. 
Since the fluctuation of the number of charged particles is a measure of the 
entropy of the system, another fluctuation observable $D-$measure of the 
net-charge provides a measure of charge fluctuations per unit entropy and is 
related to the ratio $R$ as:
\begin{eqnarray}
 D =  \langle N_{ch}\rangle\langle \delta R^2\rangle  &=& \frac{4}{\langle 
N_{ch}\rangle}\langle \delta N_+^2 + \delta N_-^2 - 2\delta N_+ \delta 
N_-\rangle \nonumber \\
&\approx& \frac{4\langle \delta Q^2\rangle}{\langle N_{ch}\rangle}
\end{eqnarray}
where $\langle \delta Q^2 \rangle$ is the variance of the net-charge with $Q = 
N_+ - N_-$ is the difference between $+$ve and $-$ve particles and 
$\langle N_{ch}\rangle = \langle N_+ + N_-\rangle$ is the average number 
of charged particles measured within the experimental acceptance. Assuming the 
quark-quark interactions to be negligible, the $D$ 
found to be approximately 4 times smaller in QGP phase as compared to HG phase. 
It has been shown that $D=$ 4 for uncorrelated pion gas, and reduces to 3 after 
taking the resonance decay into account~\cite{Jeon:2003gk}. For a QGP phase, the 
$D$ is estimated between 1.0--1.5. Hence, $D$ can be used as 
a probe to distinguish between the QGP and HG phase. Unfortunately, the 
quantity $\langle \delta Q^2\rangle/\langle N_{ch}\rangle$ depends on the 
experimental efficiency. In the experiments, the net-charge fluctuations are 
studied in terms of dynamical fluctuation measure $\nu_{(+-,dyn)}$, which is 
found to be independent of detection efficiency. The quantity $\nu_{(+-,dyn)}$ 
is a measure of the relative correlation strength of $++$, $--$ and $+-$ 
particle pairs. A positive value of $\nu_{(+-,dyn)}$ signifies the correlation 
of same charge pair, where as a negative value indicates the dominant 
contribution from correlations between opposite charges. The $\nu_{(+-,dyn)}$ is 
defined by:
\begin{eqnarray}
     \nu_{(+-,dyn)} = 
\frac {\langle N_{+}(N_{+}-1)\rangle}{\langle N_{+}\rangle^2}
&+& \frac{\langle N_{-}(N_{-}-1)\rangle}{\langle N_{-}\rangle^2} \nonumber \\
&-& 2 \frac{\langle N_{-}N_{+}\rangle}{\langle N_{-}\rangle \langle 
N_{+}\rangle}
\label{eq:nudyn}
\end{eqnarray}
where $\langle N_+\rangle$ and $\langle N_-\rangle$ are the average number of 
positively and negatively charged particles within the detector acceptance. The 
$D-$measure and $\nu_{(+-,dyn)}$ are related as $\langle N_{ch}\rangle 
\nu_{(+-,dyn)} \approx D-4$~\cite{Pruneau:2002yf}.

  \begin{figure}[htbp]
\includegraphics[width=0.5\textwidth]{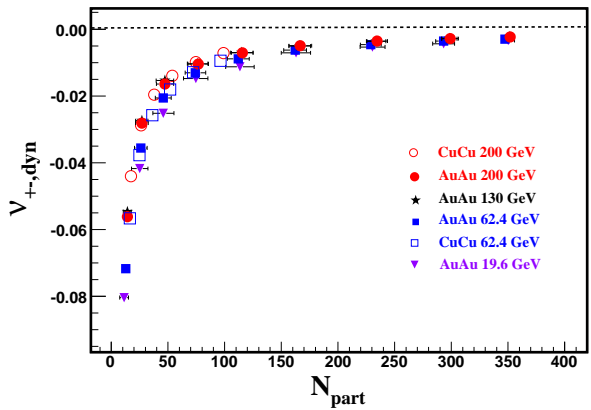}
    \caption{(Color online) Centrality ($N_{part}$) and collision energy 
dependence of dynamical net-charge fluctuations, $\nu_{(+-,dyn)}$, of particles 
produced within pseudorapidity $|\eta| <$ 0.5 in \CuCu and  \AuAu 
collisions at RHIC~\cite{Abelev:2008jg}. The combined (statistical and 
systematic) uncertainties are are within the symbol size.}
\label{fig:nudyn_npart}
  \end{figure}
Figure~\ref{fig:nudyn_npart} shows the centrality ($N_{part}$) dependence of 
dynamical net-charge fluctuations ($\nu_{(+-,dyn)}$) in \AuAu and \CuCu 
collisions at different \sqsn. The $\nu_{(+-,dyn)}$ values exhibit a monotonic 
dependence on $N_{part}$ and have small dependence on collision energy. For all 
the studied energies, the values of $\nu_{(+-,dyn)}$ are negative, which 
indicates the 
dominance of correlation of positive and negative charged particle term in 
Eq.~\ref{eq:nudyn}. The observed monotonic reduction of the magnitude
of $\nu_{(+-,dyn)}$ with increasing number of participants, arises mainly due 
to the progressive dilution of two-particle correlation when the number of 
particle sources are increased.

  \begin{figure}[htbp]
\includegraphics[width=0.5\textwidth]{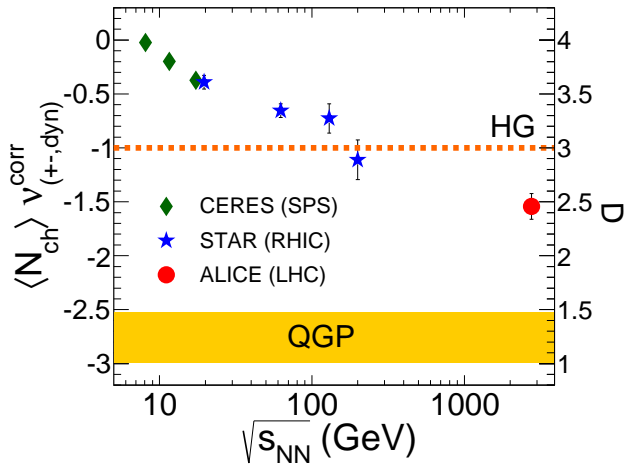}
    \caption{Energy dependence of net-charge fluctuations 
about midrapidity in central heavy-ion collisions at 
SPS~\cite{Sako:2004pw}, RHIC~\cite{Abelev:2008jg} and LHC~\cite{Abelev:2012pv} 
energies. Also shown are the expectations from a hadron resonance gas model and 
for a simple QGP picture~\cite{Jeon:2000wg}. The combined (statistical and 
systematic) errors are plotted along with the data points.}
\label{fig:nudyn_ene}
  \end{figure}
In view of prediction for critical point in the QCD phase diagram in the 
range 10$\le \sqsn \le$ 60 GeV~\cite{Abelev:2008jg,Stephanov:1999zu}, it can be 
argued that, 
the reduction of fluctuation might be larger at lower beam energies. On the 
other hand, one may also argue that, the volume of the QGP formed in \AuAu 
collisions might increase at higher beam energies leading to reduced 
fluctuations. Hence, it is important to understand whether the fluctuations may 
be found to vary with beam energy thereby indicating the production of QGP above 
a critical threshold, or with progressively increasing probability at higher 
energies. Figure~\ref{fig:nudyn_ene} shows the product of $\nu_{(+-,dyn)}$ and 
$\langle N_{ch}\rangle$ (average number of charged particles) as a function of 
collision energies for 0--5\% central collisions using the combined data 
from SPS, RHIC, and LHC 
energies~\cite{Abelev:2012pv,Abelev:2008jg,Sako:2004pw}. Also, 
the collision energy dependence of $D-$measure is shown in the same 
Fig.~\ref{fig:nudyn_ene}. It is observed that the fluctuation observable shows 
monotonic decrease in magnitude with increasing \sqsn and approaches expectation 
for a simple QGP-like scenario~\cite{Jeon:2000wg} as we move from RHIC to LHC 
energies. It has been argued that measurements in lower \sqsn (below 10 GeV) 
are dominated by baryons while at higher energies the meson and resonance 
production becomes increasingly dominant. This suggests that the change in 
dynamical net-charge fluctuations below \sqsn = 19.6 GeV might be partially due 
to this shift in particle production dominance~\cite{Abelev:2008jg,Stephanov:1998dy,
Stephanov:1999zu,Shuryak:2000pd,Aziz:2004qu,Gavin:2003cb}.
It is also argued that the differences between the fluctuation values below and 
above 19.6 GeV may result from changes in the collision dynamics and final 
state interaction 
effects~\cite{Stephanov:1999zu,Shuryak:2000pd,Aziz:2004qu,Gavin:2003cb,
Bopp:2001xc}. For the highest RHIC energy, the measured value of fluctuation 
observable is close to the HG prediction, whereas at lower energies the results 
are higher than HG value. This may be due to the fact that at highest RHIC 
energy (\sqsn = 200 GeV) the fluctuation may not be strong enough to be measured 
or because of the dilution of fluctuation during the evolution process. At LHC 
energy \sqsn = 2.76 TeV, the fluctuation observable value is significantly 
lower as compared to lower energies results. The fluctuations at the LHC energy 
might also have been diluted because of various effects, still these 
fluctuations are smaller than the theoretical expectations.
In Ref.~\cite{Jeon:2000wg}, it is shown that the $D-$measure value for hadron 
gas with resonance decay is $\simeq$ 3 and for QGP phase is $\simeq$ 
1.0$-$1.5. The measured $D-$measure value is 2.3 $\pm$ 0.21 for $\Delta\eta=$ 
1.6 at \sqsn = 2.76 
TeV. There is a clear decreasing tendency of the $D-$measure value in the HG 
phase and approaches toward QGP expectation. This may indicate that the 
fluctuations have their origin in the QGP phase~\cite{Abelev:2012pv}.

Given that several other observables already indicate that a hot and dense 
medium of color charges has been formed at RHIC and LHC energies. The 
net-charge fluctuation results suggest that either the observable  
$\nu_{(+-,dyn)}$ is not sensitive enough to QGP physics or the process of 
hadronization washed out the QGP signal for this observable. It may be also 
noted that the theoretical results do not incorporate the acceptance 
effects and dynamical evolution of the system like for example the 
dilution of the signals in the hadronization process~\cite{Singh:2013fha}.

\subsection{Fluctuation study using higher moments}
In recent years, the beam energy scan (BES) programs at SPS and RHIC have drawn 
much attention to map the quantum chromodynamics (QCD) phase diagram in terms of 
temperature ($T$) and baryon chemical potential ($\mu_B$). The location of the 
critical point can be explored by systematically varying $T$ and $\mu_B$, which 
can be experimentally achieved by varying the \sqsn of the colliding ions. 
Several theoretical models suggest that, excitation function of 
conserved numbers such as net-baryon, net-charge, and net-strangeness 
fluctuations should show a non-monotonic behavior as a possible signature of 
QCD critical end point 
(CEP)~\cite{Stephanov:2004wx,Fodor:2004nz,Stephanov:1999zu}. 
In the vicinity of the QCD critical point, these variances are proportional to 
the square of the correlation length ($\xi$), which is expected to diverge at 
the critical point~\cite{Stephanov:2011pb,cbm_book}. However, the magnitude of 
the correlation length is limited by the system size and by finite time effects 
(critical slowing down), which could be as small as 2 to 3 fm. Hence, the 
contribution to the fluctuations from the critical point might be too weak as to 
be discovered experimentally, if only the second moments are measured. 
Therefore, it has been proposed to measure higher moments of the 
fluctuations which are expected to be much more sensitive to the critical 
point~\cite{Asakawa:2009aj,Stephanov:2008qz,Hatta:2003wn}. 
The moments of the net-baryon, net-charge, and net-strangeness 
distributions are related to the $\xi$ of the system and hence these moments can 
be used to look for signals of a phase transition and critical 
point~\cite{Stephanov:1998dy,Ejiri:2005wq,Bazavov:2012vg}. 
The variance ($\sigma^2$) of these distributions 
is related to $\xi$ as $\sigma^2\sim\xi^2$, the skewness ($S$) goes as 
$\xi^{4.5}$, and the kurtosis ($\kappa$) is related as $\xi^7$. Hence it is 
proposed to study the higher moments of conserved number distribution due to 
their stronger dependence on $\xi$.

Experimentally, the net-baryon number fluctuations are not directly measured, 
as all neutral baryons are not detected by most of the experiments. Hence, 
net-baryon fluctuations are accessible via measuring the net-proton 
distributions~\cite{Aggarwal:2010wy,Adamczyk:2013dal}. The net-charge 
fluctuations are accessible by measuring the stable charged particles such as 
pions, kaons, and protons along with their 
antiparticles~\cite{Adamczyk:2014fia,Adare:2015aqk}. Similarly, the measurement
of net-kaon fluctuations acts as a proxy for net-strangeness fluctuations, 
because higher mass strange particles are not directly 
measured~\cite{Sarkar:2014wva,Thader:2016gpa}. In the following subsections we 
discuss each of the conserved number fluctuations separately.
 \begin{figure}[h]
\includegraphics[width=0.5\textwidth]{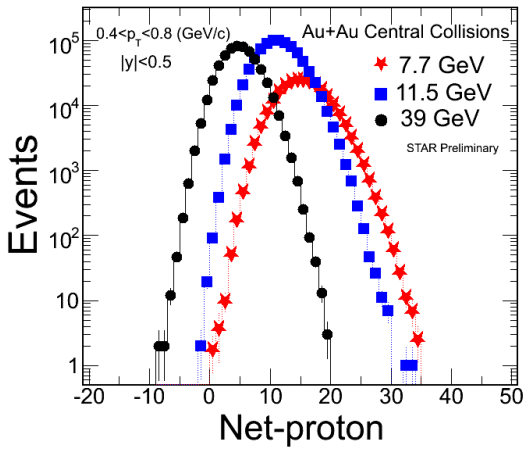}
    \caption{Uncorrected $\Delta N_p$ multiplicity distributions 
measured on an event-by-event basis in \AuAu collisions at various \sqsn for 
0--5\% collision centrality measured by STAR experiment. }
\label{fig:netp_mult}
  \end{figure}

The moments of the event-by-event experimentally measured net distributions are 
related to the different order of the cumulants of the distribution as: mean 
($M$) = $C_1$; $\sigma^2 = C_2 = 
\langle (\Delta N)^2\rangle$; $S = C_3/C_2^{3/2} = \langle (\Delta 
N)^3\rangle/\sigma^3$, and $\kappa = C_4/C_2^2 = \langle (\Delta 
N)^4\rangle/\sigma^4 - 3$, where $N$ is the multiplicity of the distribution and 
$\Delta N$ = $N-M$. Hence, the ratios of the cumulants are related to the 
moments as: $\sigma^2/M = C_2/C_1$, $S\sigma = C_3/C_2$, $\kappa\sigma^2 = 
C_4/C_2$ and $S\sigma^3/M = C_3/C_1$. Further, the ratios of moments can be 
related to the susceptibilities of $n$th order ($\chi^n$) obtained from 
the lattice QCD or from the HRG model calculations as $\sigma^2/M \sim 
\chi^{(2)}/\chi^{(1)}$, $S\sigma \sim \chi^{(3)}/\chi^{(2)}$, $\kappa\sigma^2 
\sim \chi^{(4)}/\chi^{(2)}$, and $S\sigma^3/M \sim 
\chi^{(3)}/\chi^{(1)}$~\cite{Karsch:2010ck}. One advantage of measuring the 
ratios is that the volume dependence (which is not directly measured by the 
experiment) on the experimentally measured individual cumulants cancels out to 
first order in the ratios. Hence experimentally measured quantities can be 
directly compared with the theoretical 
calculations~\cite{Gupta:2011wh,Karsch:2010ck}.

\subsubsection{Net-proton fluctuation}
Theoretical calculations have shown that net-proton fluctuations reflect the 
singularity of the charge and baryon number susceptibility as expected at the 
critical point~\cite{Hatta:2003wn}. Figure~\ref{fig:netp_mult} shows the 
typical uncorrected net-proton ($\Delta N_p$) distributions in \AuAu collisions 
for 0--5\% centrality measured on event-by-event basis. The protons and 
antiprotons are measured within $p_{T}$ = 0.4 to 0.8 GeV/$c$ and $|y|<$ 0.5. 
At lower energies, the net-proton distributions are dominated by the 
contributions from the proton distributions. The mean value of the distribution 
increases with decrease in collision energy. Different moments which describe 
the shape of the distribution are extracted from the $\Delta N_p$ 
distribution. 

\begin{figure}[htbp]
\includegraphics[width=0.5\textwidth]{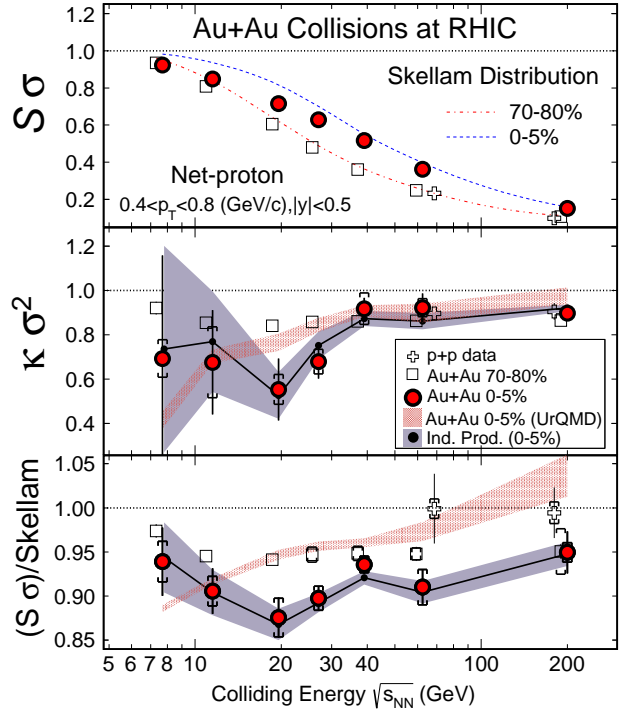}
    \caption{Collision energy and centrality dependence of 
efficiency corrected $S\sigma$ and $\kappa\sigma^2$ of net-proton distributions 
from \AuAu and $p+p$ collisions at RHIC~\cite{Adamczyk:2013dal}. 
Skellam distributions
for corresponding collision centralities are shown for $S\sigma$. Shaded 
hatched bands are the results from UrQMD. In the middle and lower 
panels, the shaded solid bands are the expectations assuming independent proton 
and antiproton production. The HRG values for $\kappa\sigma^2$ and 
$S\sigma$/Skellam are unity~\cite{Karsch:2010ck,Garg:2013ata}. The error bars 
are statistical and caps are systematic errors.}
\label{fig:netp_moments}
  \end{figure}

Figure~\ref{fig:netp_moments} shows the collision energy dependence of 
$S\sigma$ and $\kappa\sigma^2$ of net-proton distribution for 0--5\% and 
70--80\% centralities in \AuAu collisions. The statistical uncertainties are 
calculated using Delta theorem approach~\cite{Luo:2011tp}. In the hot and dense 
medium, the 
baryon chemical potential $\mu_B$ decreases with increasing collision energies, 
hence Fig.~\ref{fig:netp_moments} can be interpreted as the
$\mu_B$ dependence of the moments over the large range of $\mu_B$ (20--450 
MeV)~\cite{Adamczyk:2013dal}.
Deviations are observed for both $S\sigma$ and $\kappa\sigma^2$ from the 
Skellam and hadron resonance gas model for \sqsn $<$ 39 GeV. Maximum deviations 
from Skellam expectation are observed for \sqsn = 19.6 and 27 GeV. The 
experimental results are reasonably described by assuming independent production 
(IP) of $p$ and $\bar p$ indicating there is no apparent correlations between 
the proton and anti-protons for presented observables~\cite{Adamczyk:2013dal}. 
Therefore, one may ask, in spite of significantly correlated production due to 
baryon number, electric charge conservation and kinematical correlations of 
proton and antiprotons, why do the measured cumulants follow the independent 
production model. This has been studied by introducing the correlation between 
two independently produced distributions. It is observed that, experimentally 
measured cumulants will follow the IP model calculations even if 
the correlation coefficient is less than ∼20\%~\cite{Mishra:2015ueh}. However, 
$C_4/C_2 (= \kappa\sigma^2)$ and $C_3/C_1 (= S\sigma^3/M)$ values will follow 
the IP model for all the correlation coefficient values. The observation that 
the  experimental data can be explained by the independent production of 
particles does not rule out the existence of the critical endpoint.
  \begin{figure}
\includegraphics[width=0.5\textwidth]{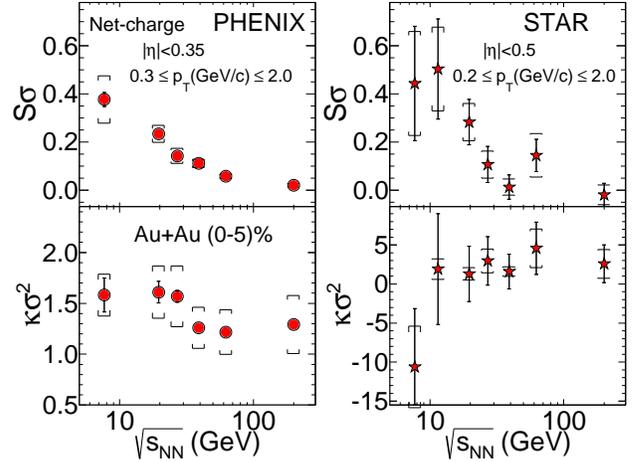}
    \caption{(Color line)The energy dependence of efficiency corrected 
$S\sigma$ and $\kappa\sigma^2$ of net-charge distributions for central 
(0-–5\%) \AuAu collisions at RHIC ~\cite{Adare:2015aqk,Adamczyk:2014fia}. The 
error bars are statistical and caps are systematic uncertainties.}
\label{fig:star_phenix_comp}
  \end{figure}
\subsubsection{Net-charge fluctuation}
The net-electric charge fluctuations are accessible through measuring 
fluctuations of stable charged particles ($\pi$, $K$, and $p$). Net-electric 
charge fluctuations are more straightforward to measure experimentally than 
net-baryon number fluctuations. As discussed before, experimentally, net-baryon 
number fluctuations are accessible only through net-proton number fluctuations, 
but in lattice QCD calculations net-baryon fluctuations are calculated instead 
of net-proton fluctuations. While net-charge fluctuations are not as sensitive 
as net-baryon fluctuations to the theoretical parameters, both measurements are 
necessary for a full understanding of the theory~\cite{Adare:2015aqk}. Several 
studies suggest that, the net-charge multiplicity distributions are better 
suited to extract the freeze-out parameters ($\mu_B$ and $T_f$) and the 
location of the CEP as they directly probe a conserved quantum 
number (electric charge)~\cite{Skokov:2011rq,Kitazawa:2011wh,Bazavov:2012vg}.

Figure~\ref{fig:star_phenix_comp} shows the collision energy dependence of 
efficiency corrected $S\sigma$ and $\kappa\sigma^2$ of the net-charge 
distributions for central 0--5\% \AuAu collisions at RHIC. The $S\sigma$ and 
$\kappa\sigma^2$ measurements from the PHENIX experiment are shown in left 
panels~\cite{Adare:2015aqk} and measurements from the STAR experiment are shown 
in right panels~\cite{Adamczyk:2014fia}. The statistical errors are calculated 
using Delta theorem method~\cite{Luo:2011tp}. Here, we would like to briefly 
discuss about the statistical error calculations in both the STAR and PHENIX 
experiments. The errors on $S\sigma$ and $\kappa\sigma^2$ are correlated and are 
dependent on both variance ($\sigma$) of the distribution and the number of 
events. As mentioned in~\cite{Luo:2011tp}, the statistical errors are more 
dependent on $\sigma$ than the number of events. Hence, experiments having 
larger multiplicity distribution will have larger errors on $S\sigma$ and 
$\kappa\sigma^2$. Since STAR experiment has larger acceptance compared to 
PHENIX, it contributes to larger statistical error. Further, STAR net-charge 
distributions have larger $\sigma$ compared to net-proton distributions, which 
contributes to the larger statistical error in the net-charge results than the 
net-proton results, although the number of analyzed events are similar. The 
$S\sigma$ values from both the 
experiments decrease with increasing \sqsn. The $\kappa\sigma^2$ values from 
PHENIX (left lower panel) remain constant and positive, between 1.0 
$<\kappa\sigma^2<$ 2.0 at all the collision energies within the statistical and 
systematic uncertainties. However, there is ∼25\% increase of $\kappa\sigma^2$ 
values at lower energies below \sqsn = 39 GeV compared to higher energies. 
Further, the $\kappa\sigma^2$ values from the STAR experiment are constant at 
all energies within uncertainties, except for \sqsn = 7.7 GeV which shows a 
negative $\kappa\sigma^2$ value. The STAR experiment reported higher weighted 
mean (2.4 $\pm$ 1.2) of $\kappa\sigma^2$ values as compared to PHENIX 
$\kappa\sigma^2$ values. It is to be noted that, results from the PHENIX 
experiment are measured within 0.3 $\le p_T$ (GeV/$c$)$\le$ 2.0 and $|\eta|\le$ 
0.35 with 2$\times \pi/2$ in azimuth, where as results from the STAR experiment 
are within 0.2 $\le p_T$ (GeV/$c$)$\le$ 2.0 and $|\eta|\le$ 0.5 with full 
azimuth (2$\pi$). Different lower $p_T$ cut may be 
responsible for about 30\% of the difference between two data 
sets~\cite{Karsch:2015zna}. More discussion on acceptance effect on higher 
moments can be found in Refs.~\cite{Garg:2013ata,Karsch:2015zna,Ling:2015yau}. 
However, net-charge results from both the experiments do not observe any 
significant non monotonic behavior in the products of moments as a function of 
collision energy.
\begin{figure}[t]
\includegraphics[width=0.45\textwidth]{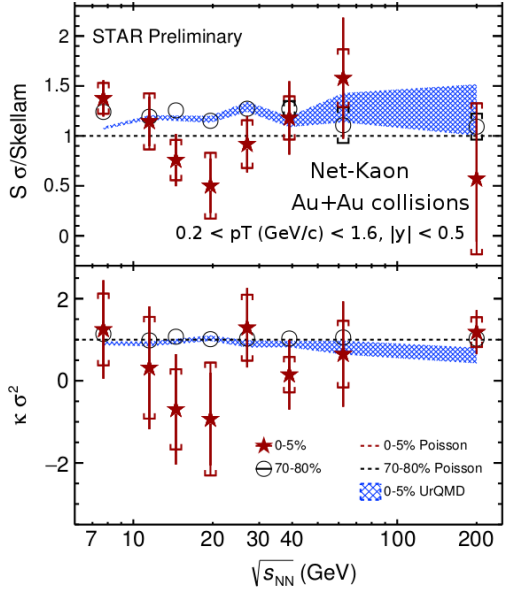}
    \caption{Energy dependence of the volume independent cumulant ratios 
of the net-kaon distributions. Showing $S\sigma$/Skellam and 
$\kappa\sigma^2$ for top 0--5\% central (red stars) and 70-80\% peripheral 
(black circles) collisions~\cite{Thader:2016gpa}. The Poisson expectations are 
denoted as dotted lines and UrQMD calculations are shown as blue bands. }
     \label{fig:netkaon}
  \end{figure}

\subsubsection{Net-Kaon fluctuation}
Experimentally, the net-strangeness fluctuations are accessible through 
measuring the net-kaon fluctuations. Figure~\ref{fig:netkaon} shows the 
preliminary results on collision energy dependence of efficiency corrected 
$S\sigma$ and $\kappa\sigma^2$ of the net-kaon distributions for 0--5\% and 
70--80\% centralities in \AuAu collisions measured by STAR experiment at 
RHIC~\cite{Thader:2016gpa}. With large 
uncertainties in the measurement, no significant deviation of the product of 
higher moments for net-kaon distributions as compared to the Poisson expectation 
has been observed at the measured energies. However, in the upcoming RHIC 
BES-II, with the upgraded detector system will help to reduce the uncertainties 
on the measurements and may find the location of the critical point.

\subsection{Extraction of freeze-out parameters using higher moments}
\begin{figure}
\includegraphics[width=0.5\textwidth]{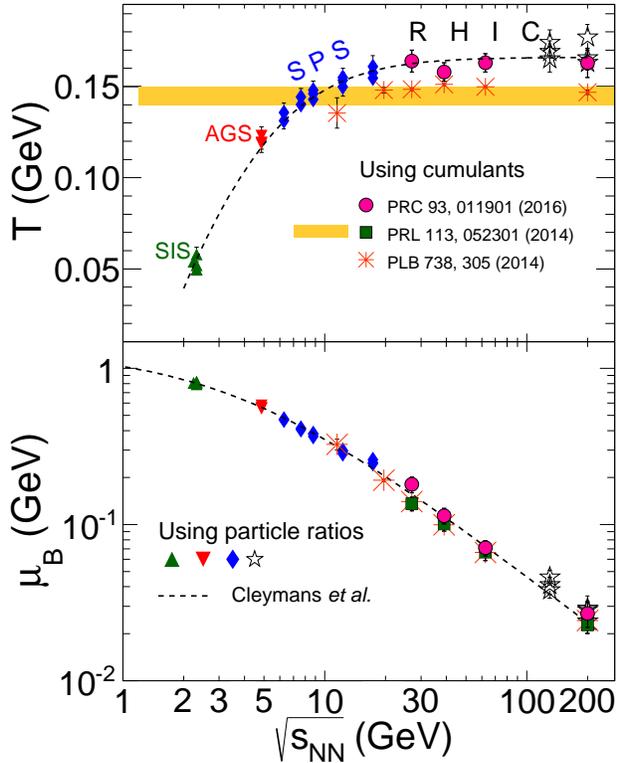}
    \caption{ The energy dependence of the chemical freeze-out 
temperature ($T_f$) and baryon chemical potential ($\mu_B$) extracted using 
lattice calculations with experimental measured 
cumulants~\cite{Borsanyi:2014ewa,Adare:2015aqk}. The freeze-out parameters 
extracted using HRG and experimental cumulants are also 
shown~\cite{Alba:2014eba}. The dashed line is the parameterization given in 
Ref.~\cite{Cleymans:2005xv}, and the SIS, AGS, SPS and RHIC data are from
Ref.~\cite{Cleymans:2005xv} and references therein.
}
\label{fig:muTvsene}     
\end{figure}

Product of higher moments can be used to extract the freeze-out parameters 
($\mu_B$ and $T_f$) of the QCD phase 
diagram~\cite{Bazavov:2012vg,Borsanyi:2014ewa,Alba:2014eba}. From the lattice 
calculations, it has been observed that, the ratio of 1$^{st}$ to 2$^{nd}$ 
cumulants ($= M/\sigma^2$) shows a strong dependence on $\mu_B$ but varies 
little with $T$. On the other hand, the ratio of 3$^{rd}$ to 1$^{st}$ cumulants 
($=S\sigma^3/M$) shows strong dependence on $T$ and has little dependence on 
$\mu_B$~\cite{Bazavov:2012vg,Borsanyi:2014ewa}. Hence, lattice calculations in 
combination with experimentally 
measured $S\sigma^{3}/M$ values at different collision energies can be used to 
extract the $T_f$, whereas the measured $M/\sigma^2$ can be used to extract the 
$\mu_B$. The collision energy dependence of freeze-out parameters $T_f$ and 
$\mu_B$ are shown in Fig.~\ref{fig:muTvsene}. Freeze-out parameters extracted 
from different methods such as using particle ratios~\cite{Cleymans:2005xv} and 
using cumulants with lattice~\cite{Adare:2015aqk} agrees very well. Further, 
the extracted $T_f$ and $\mu_B$ values using lattice calculations and 
experimental data~\cite{Adare:2015aqk} are in agreement with the thermal model 
parameterization~\cite{Cleymans:2005xv}. The freeze-out parameters extracted 
using combination of experimental data and HRG model~\cite{Alba:2014eba} 
are also shown in Fig.~\ref{fig:muTvsene}. The HRG calculations are performed 
in the same acceptance as the experiment. The extracted $\mu_B$ values using HRG 
and experimental measured cumulants are in agreement with other measurements, 
but the extracted $T_f$ are about 7--10 MeV lower as compared to the values 
extracted using lattice calculations~\cite{Alba:2014eba,Adare:2015aqk}. The 
$\mu_B$ values are consistent across different extraction procedure but there is 
some discrepancy in the value of $T_f$. One of the reason may be, $\mu_B$ are 
extracted using ratio of 1$^{st}$ and 2$^{nd}$ cumulants which are measured 
more precisely with compared to higher cumulants. In Ref.~\cite{Alba:2014eba}, 
the $\mu_B$ and $T_f$ are extracted using HRG for each \sqsn by simultaneously 
reproducing the experimentally measured ratios of the lowest-order 
susceptibilities ($M/\sigma^2$) for net-protons and net-electric charge. In 
Ref.~\cite{Borsanyi:2014ewa}, the freeze-out parameters are extracted using 
STAR experimental data~\cite{Adamczyk:2014fia}. The $\mu_B$ has been extracted 
using experimentally measured $M/\sigma^2$ and lattice calculations, where as 
due to the uncertainties on the lattice results in the low-temperature region, 
it is only possible to extract an upper value for the freeze-out temperature 
($T_f\lesssim151$ MeV). In Ref.~\cite{Adare:2015aqk}, the freeze-out parameters 
are extracted using the $M/\sigma^2$ and $S\sigma^3/M$ of net-charge from the 
PHENIX experiment~\cite{Adare:2015aqk} in combination with lattice 
calculations. Figure~\ref{fig:muTvsene} shows the direct 
combination of experimental data and theoretical model calculations to extract 
physical quantities. The consistency of the results is of fundamental importance 
to validate that the experimentally created system is close to thermal 
equilibrium at the freeze-out and can be described by lattice QCD simulations, 
at least in the light quark sector~\cite{Borsanyi:2014ewa}. 

\section{Summary}
\label{sec:summary}
In summary, we have presented a review on the experimental measurements on 
charge fluctuations over various collision energies to search for phase 
transition and location of the critical point in the QCD phase diagram. We 
reviewed the results on total charge fluctuations as a function of collision 
centrality for \sqsn = 17.3 GeV in \PbPb collisions at SPS and for \sqsn = 
62.4 and 200 GeV in \AuAu collisions at RHIC. It is observed that, the relative 
fluctuations increase from central to peripheral collisions. The fluctuation 
observables from various models are compared with the experiment data. The 
results from models show centrality independent behavior. A similar centrality 
dependence has been observed for total charge fluctuation at the SPS and RHIC 
experiments. The absence of large dynamical fluctuation in excess of the 
participant superposition model indicate that there is no evidence of critical 
behavior related to the compressibility observable.

We have also presented the results from net-charge fluctuations from SPS, RHIC 
and LHC energies. The net-charge fluctuations in terms of dynamical fluctuation 
measure $\nu_{(+-,dyn)}$ as a function of collision centrality and \sqsn are 
studied. The $\nu_{(+-,dyn)}$ increases monotonically from peripheral to 
central collisions and remains negative indicating the dominance of correlation 
of positive and negative charged particles. The product of $\nu_{(+-,dyn)}$ and 
$\langle N_{ch} \rangle$ shows a monotonic decrease with increasing 
\sqsn and approaches to the expectation from a QGP like scenario. For the 
highest RHIC energy, the measured value of fluctuation is close to the HG 
prediction, whereas at lower energies the results are higher than HG 
expectations. At LHC energy \sqsn = 2.76 TeV, the value of fluctuation 
observable is significantly lower with compared to lower energy results 
indicating at LHC energy the fluctuations have their origin in the QGP phase. 

The fluctuation of net-charge (net-proton, net-electric charge and net-kaon) 
have been measured for various \sqsn. Deviations are observed for both $S\sigma$ 
and $\kappa\sigma^2$ of net-proton distributions from the Skellam and hadron 
resonance gas model for \sqsn $<$ 39 GeV. Maximum deviations from Skellam 
expectation are observed for \sqsn = 19.6 and 27 GeV. The net-electric charge 
results from both PHENIX and STAR experiments are presented. The $S\sigma$ 
values from both the experiments decrease with increasing \sqsn. The 
$\kappa\sigma^2$ values from PHENIX remain constant and positive, between 1.0 
$<\kappa\sigma^2<$ 2.0 at all the collision energies within the statistical and 
systematic uncertainties. However, there is ∼25\% increase of $\kappa\sigma^2$ 
values at lower energies below \sqsn = 39 GeV compared to higher energies. 
The $\kappa\sigma^2$ values from STAR experiment are constant at all energies 
within uncertainties, except for \sqsn = 7.7 GeV which shows a negative 
$\kappa\sigma^2$ value. The net-electric charge and net-kaon 
results do not observe any significant non monotonic behavior as a function of 
collision energy.

To quantify the excess of net-charge fluctuations due to the critical point, 
different baseline studies have been done by different authors after considering 
various physical phenomenon. On phenomenological side, effect of resonance 
decay~\cite{Mishra:2016qyj,Nahrgang:2014fza}, proton-antiproton 
correlations~\cite{Mishra:2015ueh,Bzdak:2016sxg}, kinematic 
acceptance~\cite{Garg:2013ata,Karsch:2015zna,Ling:2015yau}, 
non-extensive systems~\cite{Mishra:2013qoa},
regeneration~\cite{Alba:2014eba,Kitazawa:2012at} and excluded 
volume models~\cite{Fu:2013gga} etc. are studied. Besides, various 
studies have been carried out using different monte carlo based models 
for example UrQMD~\cite{Westfall:2014fwa} and HIJING~\cite{Tarnowsky:2012vu} 
etc. to compare the experimental results. But it is to be 
noted that all the above mentioned physical processes contribute differently at 
various center of mass energies, therefore to find a unique ideal baseline 
estimate to compare the experimental results is still a unsettled issue. The 
freeze-out parameters are extracted using higher moments of net-charge 
distributions in combination with lattice QCD and HRG models. The extracted 
freeze-out parameters are in agreement with the thermal model parameterization. 

First results from the RHIC BES-I program are intriguing and the statistics 
collected during BES-I at RHIC are not sufficient to locate the CEP in the QCD 
phase diagram. Hence the RHIC BES-II program is proposed to have the precise  
measurements of these observables to map the QCD phase diagram. The upcoming 
RHIC BES-II program proposed to be in 2019-20, will cover the \sqsn range from 5 
to 20 GeV, which is the region of interest in the search for a critical point 
and first-order phase transition, identified by the results from BES-I and by 
model calculations. In addition, the CBM experiment at FAIR will perform a 
high-precision study of higher-order fluctuations at various beam energies in 
order to search for the elusive QCD critical point in the high net-baryon 
density region: $\mu_B\sim$  800$-$500 MeV corresponding to \sqsn = 2--4.9 GeV 
at SIS100. There are other programs at NICA to study the above physics in heavy 
ion collisions in the range \sqsn of 4--11 GeV. Hence, in the upcoming 
experimental programs at RHIC BES-II, FAIR and NICA with upgraded detector 
system will help to measure the fluctuation variable with reduced uncertainties 
on the measurements and may find the exact location of the critical point and 
investigate whole region of phase diagram in more detail.

\noindent{{\bf Conflict of interest:}}
The authors declare that there is no conflict of interest regarding the 
publication of this manuscript.


\normalsize
\begin{thebibliography}{99}
\bibitem{Stephanov:1998dy} 
  M.~A.~Stephanov, K.~Rajagopal and E.~V.~Shuryak,
  Phys.\ Rev.\ Lett.\  {\bf 81}, 4816 (1998).
  
 \bibitem{Alford:1997zt} 
  M.~G.~Alford, K.~Rajagopal and F.~Wilczek,
  Phys.\ Lett.\ B {\bf 422}, 247 (1998).
  
\bibitem{Stephanov:1996ki} 
  M.~A.~Stephanov,
  Phys.\ Rev.\ Lett.\  {\bf 76}, 4472 (1996).

  \bibitem{Aoki:2006we} 
  Y.~Aoki, G.~Endrodi, Z.~Fodor, S.~D.~Katz and K.~K.~Szabo,
  Nature {\bf 443}, 675 (2006).
%
%

\bibitem{Fukushima:2010bq} 
  K.~Fukushima and T.~Hatsuda,
  Rept.\ Prog.\ Phys.\  {\bf 74}, 014001 (2011).

\bibitem{Stephanov:2004wx} 
  M.~A.~Stephanov,
  Prog.\ Theor.\ Phys.\ Suppl.\  {\bf 153}, 139 (2004),
  Int.\ J.\ Mod.\ Phys.\ A {\bf 20}, 4387 (2005).
\bibitem{Fodor:2004nz} 
  Z.~Fodor and S.~D.~Katz,
  JHEP {\bf 0404}, 050 (2004).
\bibitem{Stephanov:1999zu} 
  M.~A.~Stephanov, K.~Rajagopal and E.~V.~Shuryak,
  Phys.\ Rev.\ D {\bf 60}, 114028 (1999).

\bibitem{Mukherjee:2015swa} 
  S.~Mukherjee, R.~Venugopalan and Y.~Yin,
  Phys.\ Rev.\ C {\bf 92}, no. 3, 034912 (2015).

\bibitem{Mukherjee:2016kyu} 
  S.~Mukherjee, R.~Venugopalan and Y.~Yin,
  Phys.\ Rev.\ Lett.\  {\bf 117}, 222301 (2016).
 

\bibitem{BES-II}
  STAR Collaboration, 
  ``Studying the Phase Diagram of QCD Matter at RHIC,''
  STAR Notes SN0598,
  https://drupal.star.bnl.gov/STAR/starnotes/public/sn0598 (2014).

\bibitem{FAIR}
  R.~Rapp {\it et al.},
  Lect.\ Notes Phys.\  {\bf 814} (2011) 335.

\bibitem{NICA}
  ``Design and construction of nuclotron-based ion collider facility (NICA) 
conceptual design report'',
  http://nica.jinr.ru/files/NICA\_CDR.pdf (2008).
  
\bibitem{JPARC}  
``J-PARC Heavy Ion Project''
  http://asrc.jaea.go.jp/soshiki/gr/hadron/jparc-hi/documents.html
\bibitem{Stodolsky:1995ds} 
  L.~Stodolsky,
  Phys.\ Rev.\ Lett.\  {\bf 75}, 1044 (1995).
\bibitem{Asakawa:2000wh} 
  M.~Asakawa, U.~W.~Heinz and B.~Muller,
  Phys.\ Rev.\ Lett.\  {\bf 85}, 2072 (2000).

\bibitem{Adams:2003st} 
  J.~Adams {\it et al.} [STAR Collaboration],
  Phys.\ Rev.\ C {\bf 68}, 044905 (2003).

  
\bibitem{Abelev:2009ai}
  B.~I.~Abelev {\it et al.}  [STAR Collaboration],
  Phys.\ Rev.\ Lett.\  {\bf 103}, 092301 (2009).
  
\bibitem{Alt:2008ab}
  C.~Alt {\it et al.}  [NA49 Collaboration],
  Phys.\ Rev.\ C {\bf 79}, 044910 (2009). 
 
\bibitem{Adcox:2002pa}
  K.~Adcox {\it et al.}  [PHENIX Collaboration],
  Phys.\ Rev.\ C {\bf 66}  024901 (2002).

\bibitem{Adler:2003xq}
  S.~S.~Adler {\it et al.}  [PHENIX Collaboration],
  Phys.\ Rev.\ Lett.\  {\bf 93}, 092301 (2004).
 

\bibitem{Jeon:2000wg} 
  S.~Jeon, V.~Koch and ,
  Phys.\ Rev.\ Lett.\  {\bf 85}, 2076 (2000).
%
\bibitem{Mrowczynski:1997kz} 
  S.~Mrowczynski,
  Phys.\ Lett.\ B {\bf 430}, 9 (1998).
\bibitem{Begun:2004pk} 
  V.~V.~Begun, M.~I.~Gorenstein, A.~P.~Kostyuk and O.~S.~Zozulya,
  Phys.\ Rev.\ C {\bf 71}, 054904 (2005).
\bibitem{Jeon:2003gk} 
  S.~Jeon and V.~Koch,
  In *Hwa, R.C. (ed.) et al.: Quark gluon plasma* 430-490
  [hep-ph/0304012].
\bibitem{Begun:2004gs} 
  V.~V.~Begun, M.~Gazdzicki, M.~I.~Gorenstein and O.~S.~Zozulya,
  Phys.\ Rev.\ C {\bf 70}, 034901 (2004).
\bibitem{Becattini:2005cc} 
  F.~Becattini, A.~Keranen, L.~Ferroni and T.~Gabbriellini,
  Phys.\ Rev.\ C {\bf 72}, 064904 (2005).

\bibitem{Garg:2015owa} 
  P.~Garg, D.~K.~Mishra, P.~K.~Netrakanti and A.~K.~Mohanty,
  Eur.\ Phys.\ J.\ A {\bf 52}, no. 2, 27 (2016)
  
\bibitem{Stanley}  
H. Stanley, {\it Introduction to Phase Transitions and Critical Phenomena} 
(Oxford, New York and Oxford) 1971.  

\bibitem{Adare:2008ns} 
  A.~Adare {\it et al.} [PHENIX Collaboration],
  Phys.\ Rev.\ C {\bf 78}, 044902 (2008).

\bibitem{Schaefer:2006ds} 
  B.~J.~Schaefer and J.~Wambach,
  Phys.\ Rev.\ D {\bf 75}, 085015 (2007).
\bibitem{Sasaki:2006ws} 
  C.~Sasaki, B.~Friman and K.~Redlich,
  Phys.\ Rev.\ D {\bf 75}, 054026 (2007).
  
\bibitem{Koch:2005vg} 
  V.~Koch, A.~Majumder and J.~Randrup,
  Phys.\ Rev.\ Lett.\  {\bf 95}, 182301 (2005).

\bibitem{Stephanov:2011pb} 
  M.~A.~Stephanov,
  Phys.\ Rev.\ Lett.\  {\bf 107}, 052301 (2011).
  
\bibitem{Asakawa:2009aj} 
  M.~Asakawa, S.~Ejiri and M.~Kitazawa,
  Phys.\ Rev.\ Lett.\  {\bf 103}, 262301 (2009).


\bibitem{Ejiri:2005wq} 
  S.~Ejiri, F.~Karsch and K.~Redlich,
  Phys.\ Lett.\ B {\bf 633}, 275 (2006).
  
\bibitem{Bazavov:2012vg} 
  A.~Bazavov, H.~T.~Ding, P.~Hegde, O.~Kaczmarek, F.~Karsch, E.~Laermann, 
  S.~Mukherjee and P.~Petreczky {\it et al.},
  Phys.\ Rev.\ Lett.\  {\bf 109}, 192302 (2012).

\bibitem{Borsanyi:2014ewa}
   S.~Borsanyi, Z.~Fodor, S.~D.~Katz, S.~Krieg, C.~Ratti and K.~K.~Szabo,
  Phys.\ Rev.\ Lett.\  {\bf 113}, 052301 (2014).
\bibitem{Alba:2014eba} 
  P.~Alba, W.~Alberico, R.~Bellwied, M.~Bluhm, V.~Mantovani Sarti, M.~Nahrgang 
  and C.~Ratti,
  Phys.\ Lett.\ B {\bf 738}, 305 (2014).

\bibitem{Aggarwal:2000aw} 
  M.~M.~Aggarwal {\it et al.} [WA98 Collaboration],
  Phys.\ Rev.\ C {\bf 64}, 011901 (2001).
%

\bibitem{Rajagopal:1993ah} 
  K.~Rajagopal and F.~Wilczek,
  Nucl.\ Phys.\ B {\bf 404}, 577 (1993).


\bibitem{Heiselberg:2000fk} 
  H.~Heiselberg,
  Phys.\ Rept.\  {\bf 351}, 161 (2001).
\bibitem{Baym:1999up} 
  G.~Baym and H.~Heiselberg,
  Phys.\ Lett.\ B {\bf 469}, 7 (1999).

\bibitem{Alt:2006jr} 
  C.~Alt {\it et al.} [NA49 Collaboration],
  Phys.\ Rev.\ C {\bf 75}, 064904 (2007).

\bibitem{Konchakovski:2005hq} 
  V.~P.~Konchakovski, S.~Haussler, M.~I.~Gorenstein, E.~L.~Bratkovskaya, 
M.~Bleicher and H.~Stoecker,
  Phys.\ Rev.\ C {\bf 73}, 034902 (2006).
 
\bibitem{Gyulassy:1994ew} 
  M.~Gyulassy and X.~N.~Wang,
  Comput.\ Phys.\ Commun.\  {\bf 83}, 307 (1994).
  
\bibitem{Cassing:1999es} 
  W.~Cassing and E.~L.~Bratkovskaya,
  Phys.\ Rept.\  {\bf 308}, 65 (1999).
 
\bibitem{Bleicher:1999xi} 
  M.~Bleicher {\it et al.},
  J.\ Phys.\ G {\bf 25}, 1859 (1999).
 
\bibitem{Werner:1993uh} 
  K.~Werner,
  Phys.\ Rept.\  {\bf 232}, 87 (1993).
  
\bibitem{Pruneau:2002yf} 
  C.~Pruneau, S.~Gavin, S.~Voloshin and ,
  Phys.\ Rev.\ C {\bf 66}, 044904 (2002).

\bibitem{Abelev:2008jg} 
  B.~I.~Abelev {\it et al.} [STAR Collaboration],
  Phys.\ Rev.\ C {\bf 79}, 024906 (2009).

\bibitem{Sako:2004pw} 
  H.~Sako {\it et al.} [CERES/NA45 Collaboration],
  J.\ Phys.\ G {\bf 30}, S1371 (2004).

\bibitem{Abelev:2012pv} 
  B.~Abelev {\it et al.} [ALICE Collaboration],
  Phys.\ Rev.\ Lett.\  {\bf 110}, no. 15, 152301 (2013).

\bibitem{Shuryak:2000pd} 
  E.~V.~Shuryak and M.~A.~Stephanov,
  Phys.\ Rev.\ C {\bf 63}, 064903 (2001).

\bibitem{Aziz:2004qu} 
  M.~A.~Aziz and S.~Gavin,
  Phys.\ Rev.\ C {\bf 70}, 034905 (2004).


\bibitem{Gavin:2003cb} 
  S.~Gavin,
  Phys.\ Rev.\ Lett.\  {\bf 92}, 162301 (2004).
\bibitem{Bopp:2001xc} 
  F.~W.~Bopp and J.~Ranft,
  Eur.\ Phys.\ J.\ C {\bf 22}, 171 (2001).

\bibitem{Singh:2013fha} 
  R.~Singh, L.~Kumar, P.~K.~Netrakanti and B.~Mohanty,
  Adv.\ High Energy Phys.\  {\bf 2013}, 761474 (2013).
  
\bibitem{cbm_book}
B. Friman et al. (eds.) Springer. 2011 The CBM Physics Book: Compressed Baryonic Matter in Laboratory Experiments vol 814 Series: Lecture Notes in Physics.


  
\bibitem{Stephanov:2008qz} 
  M.~A.~Stephanov,
  Phys.\ Rev.\ Lett.\  {\bf 102}, 032301 (2009).
\bibitem{Hatta:2003wn} 
  Y.~Hatta and M.~A.~Stephanov,
  Phys.\ Rev.\ Lett.\  {\bf 91}, 102003 (2003).
  Erratum: [Phys.\ Rev.\ Lett.\  {\bf 91}, 129901 (2003)]
\bibitem{Aggarwal:2010wy} 
  M.~M.~Aggarwal {\it et al.} [STAR Collaboration],
  Phys.\ Rev.\ Lett.\  {\bf 105}, 022302 (2010).

\bibitem{Adamczyk:2013dal} 
  L.~Adamczyk {\it et al.} [STAR Collaboration],
  Phys.\ Rev.\ Lett.\  {\bf 112}, 032302 (2014).


\bibitem{Adare:2015aqk} 
  A.~Adare {\it et al.} [PHENIX Collaboration],
  Phys.\ Rev.\ C {\bf 93}, 011901 (2016).

\bibitem{Adamczyk:2014fia} 
  L.~Adamczyk {\it et al.} [STAR Collaboration],
  Phys.\ Rev.\ Lett.\  {\bf 113}, 092301 (2014).

\bibitem{Sarkar:2014wva} 
  A.~Sarkar [STAR Collaboration],
  J.\ Phys.\ Conf.\ Ser.\  {\bf 509}, 012069 (2014).


\bibitem{Thader:2016gpa} 
  J.~Thäder [STAR Collaboration],
  arXiv:1601.00951 [nucl-ex].

\bibitem{Gupta:2011wh}
  S.~Gupta, X.~Luo, B.~Mohanty, H.~G.~Ritter and N.~Xu,
  Science {\bf 332}, 1525 (2011).
\bibitem{Karsch:2010ck} 
  F.~Karsch and K.~Redlich,
  Phys.\ Lett.\ B {\bf 695}, 136 (2011).

\bibitem{Garg:2013ata} 
  P.~Garg, D.~K.~Mishra, P.~K.~Netrakanti, B.~Mohanty, A.~K.~Mohanty, 
B.~K.~Singh and N.~Xu,
  Phys.\ Lett.\ B {\bf 726}, 691 (2013).
\bibitem{Luo:2011tp} 
  X.~Luo,
  J.\ Phys.\ G {\bf 39}, 025008 (2012).
\bibitem{Mishra:2015ueh} 
  D.~K.~Mishra, P.~Garg and P.~K.~Netrakanti,
  Phys.\ Rev.\ C {\bf 93}, 024918 (2016).
 
  
  
\bibitem{Skokov:2011rq} 
  V.~Skokov, B.~Friman and K.~Redlich,
  Phys.\ Lett.\ B {\bf 708}, 179 (2012).

\bibitem{Kitazawa:2011wh} 
  M.~Kitazawa and M.~Asakawa,
  Phys.\ Rev.\ C {\bf 85}, 021901 (2012).
\bibitem{Karsch:2015zna} 
  F.~Karsch, K.~Morita and K.~Redlich,
  Phys.\ Rev.\ C {\bf 93}, no. 3, 034907 (2016).
\bibitem{Ling:2015yau} 
  B.~Ling and M.~A.~Stephanov,
  Phys.\ Rev.\ C {\bf 93}, no. 3, 034915 (2016).
  
\bibitem{Cleymans:2005xv} 
  J.~Cleymans, H.~Oeschler, K.~Redlich and S.~Wheaton,
  Phys.\ Rev.\ C {\bf 73}, 034905 (2006).


\bibitem{Mishra:2016qyj} 
  D.~K.~Mishra, P.~Garg, P.~K.~Netrakanti and A.~K.~Mohanty,
  Phys.\ Rev.\ C {\bf 94}, 014905 (2016).

\bibitem{Nahrgang:2014fza} 
  M.~Nahrgang, M.~Bluhm, P.~Alba, R.~Bellwied and C.~Ratti,
  Eur.\ Phys.\ J.\ C {\bf 75}, no. 12, 573 (2015).
\bibitem{Bzdak:2016sxg} 
  A.~Bzdak, V.~Koch and N.~Strodthoff,
  arXiv:1607.07375 [nucl-th].
  

\bibitem{Mishra:2013qoa} 
  D.~K.~Mishra, P.~Garg, P.~K.~Netrakanti and A.~K.~Mohanty,
  J.\ Phys.\ G {\bf 42}, no. 10, 105105 (2015).

\bibitem{Kitazawa:2012at} 
  M.~Kitazawa and M.~Asakawa,
  Phys.\ Rev.\ C {\bf 86}, 024904 (2012).
  Erratum: [Phys.\ Rev.\ C {\bf 86}, 069902 (2012)].

\bibitem{Fu:2013gga} 
  J.~Fu,
  Phys.\ Lett.\ B {\bf 722}, 144 (2013).
  
\bibitem{Westfall:2014fwa} 
  G.~D.~Westfall,
  Phys.\ Rev.\ C {\bf 92}, no. 2, 024902 (2015).

\bibitem{Tarnowsky:2012vu} 
  T.~J.~Tarnowsky and G.~D.~Westfall,
  Phys.\ Lett.\ B {\bf 724}, 51 (2013).

\end{thebibliography}
\end{document}